\documentstyle[preprint,aps,epsf]{revtex}

\begin{document}

\draft

\preprint{HIP-1998-50/TH}

\title{Critical basis dependence in bounding R-parity breaking couplings 
from neutral meson mixing}

\author{Katri Huitu, Kai Puolam{\"{a}}ki and Da-Xin Zhang}

\address{Helsinki Institute of Physics, P.O.Box 9, FIN-00014
University of Helsinki, Finland}

\date{\today}

\maketitle

\begin{abstract}
Assuming  one nonzero product of two $\lambda'$-type couplings and working in
two different bases for the left-handed quark superfields,
the neutral meson mixings  are used
to bound  these  products.
We emphasize the strong basis dependence of the bounds:
in one basis many products contribute to neutral meson mixings
at tree level,
while in the other these products except one contribute
at 1-loop level only
and the GIM mechanism takes place.
Correspondingly,
these bounds differ between bases
by orders of magnitudes.
\end{abstract}

\newpage
\section{}

In the supersymmetric extensions of the standard model, 
the lepton and baryon numbers are
not necessarily conserved even at the tree-level.
These violations  of the fermion numbers can be achieved
by the breakdown of the so-called R-parity which is
defined by $R=(-1)^{(3B+L+2S)}$,
where $S$ is spin of the field \cite{dreiner}. 
An important question is then to what extent
the R-parity violating effects can be 
consistent with the present experimental data.
One class of these R-parity and lepton number violating interactions
is described by the superpotential 
\begin{equation}
W_{R\!\!\!/}=\lambda'_{ijk} L_i Q_j D^c_k ~~(i,j,k=1, 2 ~or~ 3),
\label{eq:super}
\end{equation}
where under the gauge group $SU(2)_L$
$L_i$, $Q_j$ and $D^c_k$ are superfields for the lepton doublet, 
the left-handed quark doublet and 
the right-handed quark singlet, respectively.
With these interactions special care
is needed since they contain quark mixing effects.
Another complication related to the basis is that in a general case the 
superpotential contains both bilinear and trilinear lepton number
violating interactions, among which the $R$-parity violation can be moved
by suitable field redefinitions \cite{hall}.
Here we will assume that the bilinear terms vanish.
Effects of lepton mixing with the Higgs sector have been 
extensively studied.
E.g. in \cite{DE} (lepton-Higgs)--basis independent 
measures of R-parity violation have been discussed.

In this work we will concentrate on the couplings $\lambda'_{ijk}$
in (\ref{eq:super}) and bound several products of two $\lambda'_{ijk}$'s
from neutral meson mixings.
We will follow the pioneering work by Hall and Suzuki \cite{hall}
who consider the box diagrams which
involve a scalar lepton and a weak  (or a Higgs) boson
which corresponds to the decoupling limit of the chargino or the squarks.
Recently it has been argued in \cite{rpnew}
that these box diagrams
give strict bounds on many of the products of the $\lambda'_{ijk}$'s.

We will emphasize that these bounds are sensitive to
separate rotations of the upper and lower components of the quark doublet.
We will demonstrate
the huge effect on the bounds depending on the basis chosen.
The rotations of the up- and down-type fields are also
considered in \cite{AG}, but there the strong dependence on the basis
is somewhat concealed by the assumption that only one
of the couplings is nonzero.
In \cite{AG} single couplings are bounded 
by considering neutral meson mixing through boxes involving sneutrinos
and right-handed squarks using two different bases.
In the first basis chosen, there is contribution
to the $K\bar K$ mixing, while in the second one the contribution to
the $K\bar K$ mixing is zero assuming only one nonzero coupling.
In the second basis one obtains instead bounds from $D\bar D$ mixing.
The bounds from $D\bar D$ mixing in the second basis and from 
$K\bar K$ mixing in the first basis are of the same order of 
magnitude, ${\cal{O}}(10^{-1})$.

The box contribution to the $K\bar K$
mixing induced by the R-parity violating interactions considered here is 
GIM suppressed, as will be shown.

\section{}

Before calculating the diagrams directly,
we specify our conventions for the quark superfields.
We will work in a basis in which the right-handed quarks 
in the superfield $D^c_k$'s are the mass eigenstates.
The left-handed quarks in $Q_j$ are weak eigenstates
which are related to the mass eigenstates by the general rotations
\begin{equation}
u_L^{I}=V_{uL}^{\dagger} u_L,~
d_L^{I}=V_{dL}^{\dagger} d_L,\label{basis}
\end{equation}
and the quark mixing matrix is $V\equiv V_{uL}V_{dL}^\dagger$.
At this stage we must specify the basis in which we are working,
if we assume that only one product of two $\lambda'$'s is nonzero.
Although there is no difference in physics between these bases, 
assuming that only one coupling or one product of the couplings is nonzero
corresponds to different assumptions made in different bases
in bounding the coupling(s).
Instead of discussing the most general basis,
we choose two different bases:
{\it basis I} with $V_{uL}=1$ and {\it basis II} with $V_{dL}=1$.
In the {\it basis I}, all the products of the form of
$\lambda'_{im1}\lambda'_{in2}$ contribute at tree-level to $K\bar K$ mixing.
In the {\it basis II}, only 
$\lambda'_{i21}\lambda'_{i12}$
contributes at tree-level,
while all the other $\lambda'_{im1}\lambda'_{in2}$'s
contribute at 1-loop level to the same process.
Similar observation follows 
in the cases of $B\bar B$ and $B_s\bar B_s$ mixings.
When assuming that only one product is nonzero,
the difference between these two bases is critical.

\section{}

First we take the {\it basis I}: $V_{uL}=1$ and thus $V= V_{dL}^\dagger$.
We take  $K\bar K$ mixing as an example.
In this basis,
the tree level FCNC process can be induced by any product of
the form $\lambda'_{im1} \lambda'_{in2}$'s,
and we will denote the coupling  $\lambda'_{imn}$ 
in this basis by $\bar\lambda'_{imn}$.
The tree level neutral couplings induced by $\bar\lambda'_{imn}$
can be read from (\ref{eq:super}) and (\ref{basis}),
and thus $K\bar K$ mixing is induced at tree level by any
product of  $\bar\lambda'_{im1}\bar \lambda'_{in2}$ as
\begin{eqnarray}
{\cal H}_{tree}(\Delta S=2)=\bar\lambda'_{im2}\bar \lambda^{\prime *}_{in1}
V_{m1}V_{n2}^* \displaystyle\frac{1}{m_{\tilde \nu}^2}
\bar s_R d_L\bar s_L d_R.
\label{hbar}
\end{eqnarray}
Consequently,
if we assume that only one of these products is nonzero,
the bounds on these products are very strong,
and they are given in Table 1.

\section{}

Next we take the {\it basis II}: $V_{dL}=1$ and thus $V= V_{uL}$.
We start with $K\bar K$ mixing.
Assuming that only one $\lambda'$ product contributes,
the box diagrams give\footnote{The complete calculation is given elsewhere.}
\begin{eqnarray}
{\cal H}^{mn}_{box}(\Delta S=2)
&=&\frac {g_2^2}{32 \pi^2 m_W^2}
\lambda'_{im2}\lambda_{in1}^{\prime *}
\bar s_R d_L\bar s_L d_R
\sum_{X=W,G,H}\sum_{k,h=1}^3 V_{k1}V_{km}^*V_{hn}V_{h2}^*
F_X^{kh}\nonumber\\
&&\hspace{-0.5cm} 
=\frac {g_2^2}{32 \pi^2 m_W^2}\lambda'_{im2}\lambda_{in1}^{\prime *}
\bar s_R d_L\bar s_L d_R
\sum_{X=W,G,H}\nonumber\\
&&
\{V_{3n}V_{32}^*[V_{31}V_{3m}^*(F_X^{33}-F_X^{31}-F_X^{13}+F_X^{11})
                 +V_{21}V_{2m}^*(F_X^{23}-F_X^{21}-F_X^{13}+F_X^{11})]
\nonumber\\
&& +V_{2n}V_{22}^*[V_{31}V_{3m}^*(F_X^{32}-F_X^{31}-F_X^{12}+F_X^{11})
                 +V_{21}V_{2m}^*(F_X^{22}-F_X^{21}-F_X^{12}+F_X^{11})]
\nonumber\\
&& +\delta_{n2}   [V_{31}V_{3m}^*(F_X^{31}-F_X^{11})
                 +V_{21}V_{2m}^*(F_X^{21}-F_X^{11})]\nonumber\\
&& +\delta_{m1}   [V_{3n}V_{32}^*(F_X^{13}-F_X^{11})
                 +V_{2n}V_{22}^*(F_X^{12}-F_X^{11})]\nonumber\\
&& +\delta_{n2} \delta_{m1} F_X^{11}\},
\label{KGIM}
\end{eqnarray}
where the $F_X^{ij}$'s are the contributions from the box diagram
$X- {\tilde e} - u_i - u_j ~(X=W, G, H)$.
In the second equality of (\ref{KGIM}),
we have used the unitarity condition $V V^\dagger= V^\dagger V =1$.
Note that the last term of the second equality
has a tree-level correspondence given in \cite{chou}, 
which differs from 
the contribution from $\bar\lambda'_{i21}\bar \lambda'_{i12}$
in {\it basis I} by only a factor of $1/(V_{22}V_{11}^*)\sim 1$.
The effect of the box diagram on $\lambda'_{i21} \lambda'_{i12}$
is  thus  negligible\footnote{Our bound for the
product $\lambda^\prime_{i13}\lambda^\prime_{i31}$ (given in Table I)
from $B\bar B$ mixing,
differs from $3.3\times 10^{-8}$ given in [4],
and $8\times 10^{-8}$ given in [3] from the same process.}.
We will discard this term below.
Analytically,
\begin{eqnarray}
F^{21}-F^{11}&=& c \left( \frac{\ln c}{e}+\frac{\ln e}{e(e-1)}\right)  
\nonumber \\ 
F^{31}-F^{11}&=&t \left( \frac{\ln t}{(t-e)(t-1)}+\frac{\ln e}{(t-e)(1-e)}
\right)  
\nonumber \\ 
(F^{22}-F^{21})-(F^{12}-F^{11})&=&\frac {c}{e}   
\nonumber\\
(F^{32}-F^{31})-(F^{12}-F^{11})&=& c \left[- \frac{\ln c}{e}
+\frac{(1+t)\ln t}{(t-e)(t-1)} -\frac{t (1+e)\ln e}{e (t-e)(t-1)}\right.    
\nonumber\\
&&\hspace{-0.8cm}  
\left. -t \cot^2 \beta \left( -\frac{\ln t}{(t-e)(t-h)}
+\frac{\ln e}{(t-e)(e-h)}
- \frac{\ln h}{(t-h)(e-h)} \right) \right]   \nonumber\\
(F^{33}-F^{31})-(F^{13}-F^{11})&=& t \left[ \frac{1+t}{(t-e)(t-1)}
+ \frac{t (1+2e-2t-t^2) \ln t}{(t-e)^2(t-1)^2}
+\frac{t(1+e) \ln e}{(t-e)^2(e-1)}   \right.
\nonumber \\
&&\hspace{-4.5cm}  
-t \left. \cot^2 \beta \left( -\frac 1{(t-e)(t-h)}
+\frac{(t^2-eh) \ln t}{(t-e)^2 (t-h)^2}
-\frac{e \ln e}{(t-e)^2 (e-h)}+\frac{h \ln h}{(t-h)^2 (e-h)} \right)\right],
\label{abox}
\end{eqnarray}
where we have used the following dimensionless quantities:
\begin{eqnarray}
c=m_c^2/m_W^2, t=m_t^2/m_W^2, e=m_{\tilde{e}_n}^2/m_W^2,
h=m_{H^\pm}^2/m_W^2, \tan\beta =  \langle H_u^0 \rangle/\langle H_d^0
\rangle .
\end{eqnarray}
We have summed over the $W, G, H$ contributions in eqn. (\ref{abox}).

Eqs. (\ref{KGIM}-\ref{abox}) show explicitly 
that the GIM cancellations take place
also in the FCNC processes 
induced by the R-parity violating interactions.
The $(F_X^{33}-F_X^{31})-(F_X^{13}-F_X^{11})$ 
and $F_X^{31}-F_X^{11}$ terms depend
only on the large masses $m_t$, $m_W$, etc.,
but these terms are suppressed by the small entries of the CKM matrix $V$.
The other contributions are suppressed by the small 
$c={m_c^2}/{m_W^2}$.
This is a general feature of the chosen basis.
Consequently,
those products of $\lambda'$'s which have large CKM factors
for the contributions and which are free from the mass suppression
will get strong bounds from $K-\bar K$ mixing.
In the case of $B_d-\bar B_d$ mixing,
the mass suppressed terms are usually related to the small
CKM matrix elements and are thus less relevant.

The amplitude for $K-\bar K$ mixing can
be calculated using vacuum insertion method and equation of motions
for the quarks \cite{mix}.
The relevant matrix element turns
out to be 
$\langle K | ({\overline{d}} P_R s ) ({\overline{d}} P_L s) | 
\bar K \rangle = B_K f_K^2 M_K^2 
(\frac{1}{2} M_K^2/(M_s+M_d)^2+\frac{1}{12})$,
and  the mass splitting is 
$\Delta m_K={\text{Re}}(\langle K |{\cal{H}}| \bar K \rangle )/M_K$. 
The corresponding amplitude for $B_d-\bar B_d$ mixing is 
$\langle B_d | ({\overline{d}} P_R b ) ({\overline{d}} P_L b ) | 
\bar B_d \rangle = B_{B_d}f_{B_d}^2 M_{B_d}^2 
(\frac{1}{2} M_{B_d}^2/(M_b+M_d)^2+\frac{1}{12})$,
where $f_K=0.15$ GeV and $f_{B_d}=0.2$ GeV.
We will also use $B_K=0.75$, $B_{B_d}=1$,
$M_s+M_d=0.175$ GeV and $M_b+M_d=4.8$ GeV.

Our constraints in the {\it basis II} are summarized in Table 2,
where we use $\Delta M_K=3.49 \times 10^{-12}$ MeV
and $\Delta M_{B_d}=0.474$ ps$^{-1}$ \cite{PDG}.
We also take the charm quark
mass as $m_c=1.3$ GeV, the top mass as $m_t=175$ GeV,
and  three CKM angles to be $\sin \theta_{12}=0.219$, $\sin
\theta_{23}=0.041$ and $\sin \theta_{13}=0.0035$ \cite{PDG}. 
In our numerical
results we take the slepton masses to be
$m_{\tilde{\nu}_n}=m_{\tilde{e}_n}=100$ GeV. 
We use two representative
values of $\tan \beta$:
$\tan \beta=1$ and $\tan \beta=50$. We are left with two
free parameters: the charged Higgs boson mass $m_{H^\pm}$ and the
complex phase $\delta$ of the CKM matrix.

Although the mixing in the case of {\it basis II} is in the up-quark 
sector, note that the $D\bar D$
mixing cannot give tree level bounds, since in the superpotential 
the interactions with $\lambda '$-type couplings
always contain down-type quarks.

\section{}

We have investigated two different bases of the 
$\lambda^\prime$-type R-parity breaking model.
A comparision of the bounds on $\bar \lambda' \bar\lambda'$
in Table I and $\lambda'\lambda'$ in Table II
is impressive showing the importance of specifying
the basis in which the bounds on couplings are found.
Assuming only one nonzero product of the $\lambda'$'s
but working in different bases,
the bounds can arise from either tree or 1-loop diagrams,
and the numbers can differ between the bases by orders of magnitudes.

This work is partially
supported by the Academy of Finland (no. 37599).

\newpage

\begin{table}
\caption{
Bounds on all the products $\bar\lambda'_{nj1} \bar\lambda'_{nk2}$
from $\Delta M_K$, 
and on $\bar\lambda'_{nj1} \bar\lambda'_{nk3}$
from $\Delta M_{B_d}$.
Numbers are given for $m_{\tilde \nu}=100$ GeV.
}
\begin{tabular}{lc|lc|lc}
$(n11)(n12)$&$4.5\times 10^{-9}$&
$(n11)(n22)$&$2.0\times 10^{-8}$&
$(n11)(n32)$&$8.1\times 10^{-7}$\\
$(n21)(n12)$&$9.8\times 10^{-10}$&
$(n21)(n22)$&$4.5\times 10^{-9}$&
$(n21)(n32)$&$1.7\times 10^{-7}$\\
$(n31)(n12)$&$2.3\times 10^{-8}$&
$(n31)(n22)$&$1.1\times 10^{-7}$&
$(n31)(n32)$&$4.3\times 10^{-6}$\\
\tableline
$(n11)(n13)$&$6.0\times 10^{-6}$
&$(n11)(n23)$&$2.7\times 10^{-5}$
&$(n11)(n33)$&$1.1\times 10^{-3}$\\
$(n21)(n13)$&$5.2\times 10^{-7}$
&$(n21)(n23)$&$2.4\times 10^{-6}$
&$(n21)(n33)$&$1.0\times 10^{-4}$\\
$(n31)(n13)$&$2.1\times 10^{-8}$
&$(n31)(n23)$&$1.0\times 10^{-7}$
&$(n31)(n33)$&$3.8\times 10^{-6}$\\
\end{tabular}
\end{table}

\begin{table}
\caption{New bounds on the products $\lambda'_{nj1} \lambda'_{nk2}$
from $\Delta M_K$, 
and on $\lambda'_{nj1} \lambda'_{nk3}$
from $\Delta M_{B_d}$.
These bounds are stronger than those given in the literature.
Numbers are given for $m_{\tilde e}=100$ GeV, $m_{H^+}=100$ GeV and 
$\tan\beta= 1$ and $50$,
separated by slash.
Numbers in the parenthesis are for $m_{H^+}=1000$ GeV.
}
\begin{tabular}{lccc}
 & $\delta=0$ & $\delta=\pi/2$ & $\delta=\pi$ \\ 
$(n11)(n12)$ & $1.2 \times 10^{-3}$ &  $1.0 \times 10^{-3}$ &  
$8.8 \times 10^{-4}$ \\ 
$(n21)(n22)$ & $1.3 \times 10^{-3}$ &  $1.1 \times 10^{-3}$ &  
$9.5 \times 10^{-4}$ \\ 
$(n21)(n32)$ & $1.5 \times 10^{-4}$ &  $9.5 \times 10^{-5}$ &  
$6.9 \times 10^{-5}$ \\ 
$(n31)(n12)$ & $2.1 \times 10^{-5}$ &  $2.1 \times 10^{-5}$ &  
$2.3 \times 10^{-5}$ \\ 
$(n31)(n32)$ & $0.0040/0.027$ &  $0.0026/0.018$ &  $0.0019/0.013$  \\ 
	   &  ($0.022/0.027$)&  ($0.014/0.018$)& ($0.010/0.013$) \\ 
\tableline
$(n11)(n13)$ & $0.0035$ &  $0.0021$ &  $0.0015$ \\ 
$(n21)(n13)$ & $4.7 \times 10^{-4}$ &  $4.8 \times 10^{-4}$ &  
$4.9 \times 10^{-4}$ \\ 
$(n21)(n33)$ & $0.092/0.62$ &  $0.058/0.39$ &  $0.043/0.29$  \\ 
	   &  ($0.51/0.63$)&  ($0.32/0.40$)& ($0.24/0.30$) \\ 
$(n31)(n23)$ & $0.036/0.059$ &  $0.025/0.041$ &  $0.019/0.031$ \\ 
           &  ($0.058/0.059$)& ($0.040/0.041$)& ($0.031/0.031$)\\ 
$(n31)(n33)$ & $0.0019/0.0031$ &  $0.0011/0.0019$ &  $0.00081/0.0013$  \\ 
	   &  ($0.0030/0.0031$)&  ($0.0019/0.0019$)& ($0.0013/0.0013$) \\ 
\end{tabular}
\end{table}

\end{document}